\documentclass[nofootinbib,twocolumn,aps,prd,amsmath,superscriptaddress]{revtex4-1}
\usepackage{setspace}
\usepackage{color}
\usepackage{fancyhdr}
\usepackage[ansinew]{inputenc}
\usepackage{amssymb}
\usepackage{amsmath}
\usepackage{cancel}
\usepackage{graphicx}
\usepackage{float}
\usepackage{subfigure}
\usepackage{cancel}
\usepackage{tabularx}
\usepackage{cleveref}

\newcommand{\rom}[1]{\MakeUppercase{\romannumeral #1}}
\begin{document}
%===============================================================================
\title{Signals in the tidal deformability for phase transitions in compact stars with constraints from GW170817}
%===============================================================================
%==============================-Autoren-========================================
%===============================================================================
\author{Jan-Erik Christian}
\email{christian@astro.uni-frankfurt.de}
\affiliation{Institut f\"ur Theoretische Physik, Goethe Universit\"at Frankfurt, 
	Max von Laue Strasse 1, D-60438 Frankfurt, Germany}

\author{Andreas Zacchi}
\email{zacchi@astro.uni-frankfurt.de}
\affiliation{Institut f\"ur Theoretische Physik, Goethe Universit\"at Frankfurt, 
Max von Laue Strasse 1, D-60438 Frankfurt, Germany}

\author{J\"urgen Schaffner-Bielich}
\email{schaffner@astro.uni-frankfurt.de}
\affiliation{Institut f\"ur Theoretische Physik, Goethe Universit\"at Frankfurt, 
Max von Laue Strasse 1, D-60438 Frankfurt, Germany}
%============================================================================
\date{\today}
%============================================================================

  %==========================================================================
  \begin{abstract}
  %=============================================================================
We compute the tidal deformabilities for neutron star merger for
equations of state with a strong first order phase transition producing
a new separate branch in the mass-radius diagram.  A case is found where
all three possible pairs of combinations between these two neutron star
branches are present for the total mass of $M=2.7M_\odot$ of the
observed merger event GW170817. It is demonstrated that the plot of the
two tidal deformabilities $\Lambda_1$ and $\Lambda_2$ of the binary
neutron star can show up to three separate branches. We propose that the
future detections of neutron star merger events with the same value for
$\Lambda_1$ but different values of $\Lambda_2$ serve as a signal for
the existence of a strong first order phase transition in neutron star
matter.
  %=============================================================================
  \end{abstract}
\maketitle
%============================================================================
%-=====================   The sections   =======================================
%============================================================================
  \section{Introduction}\label{incanus}
%=============================================================================
% It has been speculated for half a century, that a 
% class of compact stars may arise in form of hybrid stars, whose core consists
% of quark matter due to a phase transition \cite{Alford:2017qgh,Zacchi:2015oma,Zacchi:2015lwa,Zacchi:2016tjw,Christian:2017jni}.  
It has been speculated for years that compact stars may form a separate branch (third family) 
in the mass-radius relation of compact stars, separated from ordinary neutron stars by an
instability region analogous to the one existing between white dwarfs and neutron stars.  
These so called twin stars have been discussed in the literature a long time  \cite{1981JPhA...14L.471K,Glendenning:1998ag,Schertler:2000xq,SchaffnerBielich:2002ki,Zdunik:2012dj,Alford:2015dpa,Blaschke:2015uva,Zacchi:2016tjw,Alford:2017qgh,Christian:2017jni} 
and it is widely believed that a phase transition is responsible for the 
appearance of a second stable branch in the mass radius relation. \\
At large densities hadronic matter is expected to undergo two phase transitions.  
The first one deconfines hadrons to quarks and gluons. 
The second one restores chiral symmetry. Yet it is an unsettled issue 
whether these transitions are real phase transitions or
crossover transitions \cite{Alford:2017qgh,Alford:2015dpa,Blaschke:2014via,Zacchi:2015oma}. 
Even the possibility of the existence of pure quark stars is not ruled out yet \cite{Ivanenko:1965dg,Itoh:1970uw,Bodmer:1971we,Haensel:1986qb,Alcock:1986hz,Fraga:2001xc,Zacchi:2015lwa}.
The radius measurement carried out by the NICER experiment \cite{2014SPIE.9144E..20A} 
may elucidate the issue on these phase transitions, because 
the discovery of two stars with the same
masses but different radii could be indeed a signal 
of a phase transition in dense matter.\\
The equation of state (EoS) of dense matter constructing stellar models is not only a key ingredient in modeling compact star formation,
but also sensitive to the gravitational wave signal from mergers of binary neutron stars \cite{TheLIGOScientific:2017qsa,Annala:2017llu,Bauswein:2017vtn,Paschalidis:2017qmb}. 
Apart from the two solar mass limit of the %recently 
%discovered 
pulsars PSR J1614-2230 and of PSR J0348+0432 \cite{Demorest:2010bx,Antoniadis:2013pzd,Fonseca:2016tux}, 
new stringent limits on the EoSs came 
from the LIGO/Virgo detection of gravitational waves originating from a neutron-star merger,
GW170817, which has provided limits on the tidal deformabilities of the stars 
involved in the collision \cite{TheLIGOScientific:2017qsa,Most:2018hfd,Rezzolla:2017aly}. 
It is possible to use the tidal Love number as a criterion for the validity of EoSs \cite{Annala:2017llu,Tews:2018iwm,Abbott:2018exr,De:2018uhw}.

In this work we explore the tidal deformabilities of the twin star solutions 
obtained with constant speed of sound parametrized EoSs (see Alford et al. \cite{Alford:2013aca}) as studied in 
\cite{Christian:2017jni}. Different values of the critical pressure and the energy density 
jump allow for sorting the twin star solutions into four categories.\\
The corresponding tidal deformabilities of the four categories for twin stars studied in \cite{Christian:2017jni} allow for a further constraining of EoSs with a sharp phase transition.\\
We find that the merger of a neutron-neutron, a neutron-hybrid and a hybrid-hybrid star is possible, where we define the hybrid star as a compact star with a hadronic crust and a quark matter core. The resulting 
$\Lambda_1-\Lambda_2$ plot has three distinct branches. For larger total mass the $\Lambda_1-\Lambda_1$ band shifts to lower values generally, because the merging stars are more compact for higher masses. 
%=============================================================================
%=========================================================================
   \section{Theoretical framework}\label{micheldelving}
%==========================================================================
%==============================
\subsection{Phase transition and equation of state} 
%==========================================================================
Based on the assumption that the phase transition at high baryonic 
densities is of first order%\cite{Fodor:2001pe,Fodor:2007ue,Alvarez-Castillo:2013cxa}
, the phase transition from hadronic matter to quark matter in our approach is modeled via a 
Maxwell construction.\\ %\cite{Glendenning:1992vb,Bhatta04,Bhattacharyya:2009fg}.\\ 
For the EoS for the hadronic phase we take the widely used model DD2 by Typel et al. \cite{Typel:2009sy}. For the stellar quark matter core a constant speed of sound parametrized EoS, introduced by Alford et al. \cite{Alford:2013aca,Alford:2015dpa} (see for an earlier version\cite{Zdunik:2012dj}) is utilized. This EoS is given as 
 \begin{equation}
  \epsilon(p) =
 	\begin{cases} 
 		\epsilon_{DD2}(p)	&  p < p_{trans}\\
 		\epsilon_{DD2}(p_{trans})+\Delta\epsilon + c_{QM}^{-2}(p-p_{trans})	& p > p_{trans}\\
 	\end{cases}
 \end{equation} 
where $\epsilon$ is the energy density, $p$ the pressure and $p_{trans}$ the transitional pressure. %The energy density counterpart is $\epsilon_{trans}=\epsilon_{DD2}(p_{trans})$.
The discontinuity in energy density at the transition is $\Delta\epsilon$. %Throughout this work we work in natural units ($\hbar=G=c=1$) and 
For the speed of sound in the stars core, a value of $c_{QM}^2=1$ is assumed (using natural units).This stiffest possible EoS allows for the broadest range of twin star solutions and acts as an upper bound, so that $c_s^2=1$ in the quark matter phase allows for the smallest radii possible \cite{Alford:2015dpa}.
For more details see the references above and the references in \cite{Christian:2017jni}.
%==========================================================================
\subsection{Classification by mass}\label{Tevildo}
%==========================================================================
This subsection is a brief compendium on the categories of twin stars for the readers convenience. In the following the first and second maximum will refer to the maximum of the hadronic branch and the maximum of the hybrid star branch in a twin star mass-radius relation respectively.
In \cite{Christian:2017jni} we showed that for nonrotating compact stars the value of $\Delta\epsilon$ has virtually no influence on the mass at the first maximum but affects the second maximum in the mass radius relation by determining the position of the hybrid star branch in the MR-relation.
Likewise, it is possible to assign a specific mass from the first maximum to a specific $p_{trans}$. The value of $p_{trans}$ has no influence on the shape of the first branch, only on its maximum, however it is crucial for the shape of the second branch. The twin star categories are defined as follows:
\begin{itemize}
 \item [\textbf{\rom{1}:}] Both maxima exceed $2M_\odot$, which implies high values of $p_{trans}$. This category features the heaviest twin stars with $M=2.24M_\odot$ and the second branch is nearly flat.  
 \item [\textbf{\rom{2}:}] The mass of the first maximum exceeds $2M_\odot$, again implying high $p_{trans}$, whereas stars from the second branch do not reach $2M_\odot$. As in Category I the second branch is nearly flat.
 \item [\textbf{\rom{3}:}] Defined by demanding that the mass of the first maximum is in the range $2M_\odot \geq M_{max_1} \geq 1M_\odot$ and the second maximum exceeds $2M_\odot$. The second branch rises much steeper in mass compared to the previous categories, since the transitional pressure is lower.
 \item [\textbf{\rom{4}:}] The first maximum is below one solar mass and the second maximum exceeds $2M_\odot$. EoSs generating a mass radius relation in this category have radii which are nearly constant for increasing mass.  
\end{itemize}

In this article we will examine five specific EoSs from \cite{Christian:2017jni} in order to illustrate the general cases of binary neutron star merger (BNSM) scenarios. The defining parameters of these EoSs are listed in table \ref{Annatar} and the corresponding mass-radius relations are shown in Fig.\ref{MR_Plot}. 
%%%%%%%%%%%%%%%%%%%%%%%%%%%%%%%%%%%%%%%%%%%%%%%%%%%%%%%%%%%%%%%%%%%%%%%%%%%
\begin{table}[ht]
	\begin{center}
		\begin{tabular}{|c|c|c|c|c|}
			\hline
			% \multicolumn{1}{|c|}{J=32~MeV} & \multicolumn{2}{c}{Just the physical ones...} & \multicolumn{2}{c|}{} \\
			% \hline
			\hspace{.2cm} Case \hspace{.2cm} & \hspace{.01cm} Category
			\hspace{.1cm} & \hspace{.2cm} $p_{trans}$ \hspace{.2cm} & \hspace{.1cm}
			$\Delta \epsilon$ \hspace{.1cm} & \hspace{.1cm}
			$M_{twin}$ ($M_\odot$) \hspace{.1cm}  \\
			\cline{1-5}
			% \cline{2-5}
			%  0.50 & 0.60 & 12.7990 & 0.0449062    & \\
			% \cline{2-6}
			CII & \rom{2} & 118 & 718  & 1.66  \\ % & 1.66 ?
			%   Cs: 389.033 Cv: 264.349
			CIIIa & \rom{3} & 64 & 354 & 1.59 - 1.65 \\ % &  1.55-1.65 ?
			%  Cs: 352.269 Cv: 232.267
			CIIIb & \rom{3} & 40 & 350 & 1.28 - 1.346   \\ %  1.3-1.34?
			% Cs: 317.031 Cv: 199.12
			CIIIs & \rom{3} & 43.5 & 350 & 1.39 - 1.32  \\ % 1,35
			CIV & \rom{4} & 7  & 400 & -   % keine
			
			%			M_Twin
			%%CII: 		1.659 - 1.662
			%%CIIIa:	1.648 -  1.585
			%%CIIIb:	1.346 -  1.279
			%%CIIIs:	 1.387 - 1.32
			%%CIV:		/
			
			% \cline{2-6}
			% \cline{2-6}
			\\ \hline
			% \hline
		\end{tabular}
		\caption{These are the cases examined in \cref{Manwe}, with choices based on \cite{Christian:2017jni}. The values of $p_{trans}$ and $\Delta\epsilon$ are displayed in units of $\mathrm{MeV/fm^3}$.}
		\label{Annatar}
	\end{center}
\end{table}
%&&&&&&&&&&&&&&&&&&&&&&&&&&&&&&&&&&&&&&&&&&&&&&&&&&&&&&&&&&&&&&&&&&&&&&&&&&&&&&&&&&&&&&&&&&&&&&&&&&&&&&&&&
Case CII is a typical example of category \rom{2}. Case CIIIa is an EoS with its twin star mass above $1.35M_\odot$, which is roughly $0.5M_{total}$ measured by LIGO for GW170817. The case CIIIb contains twin stars below this mass and CIIIs denotes a special case where the twin stars are located exactly at $1.35M_\odot$. The EoS CIV is an extreme example of category \rom{4} and has not only the lowest value of $p_{trans}$ of this category but also a very high $\Delta\epsilon$. The twin star masses $M_{twin}$ are possible mass ranges of all twin star pairs in the respective EoS. 
%============================================

%==========================================================================
\subsection{Tidal deformability}\label{gandalf}
%==========================================================================
The observation of the event GW170817 detected by the LIGO and Virgo
observatories \cite{TheLIGOScientific:2017qsa} can be used to constrain
the EoSs of compact stars, because  
GW170817 sets limits on the tidal deformability during the inspiral
phase of a neutron star merger.
The tidal deformability $\lambda$ measures one stars quadrupole
deformation $Q_{ij}$ in response to the companions tidal field
$\mathcal{E}_{ij}$ \cite{Hinderer:2007mb,Hinderer:2009ca}.
\begin{equation}
Q_{ij}=-\lambda \mathcal{E}_{ij} .
\end{equation}
Here $\lambda$ is the ratio of the induced quadrupole moment to the
perturbing tidal field, i.e.
$$\lambda=\rm{\frac{induced \,\, quadrupole}{perturbing \,\, tidal \,\,
		field}}$$
$\lambda$ itself, which also depends on the EoS, is related to the stars quadrupolar tidal Love number
$k_2$ via 
\begin{equation}
k_2=\frac{3}{2}\lambda R^{-5} \, ,
\end{equation}
R being the radius of the star.

Once $k_2$ is known, the dimensionless tidal deformability $\Lambda$ can be computed as in \cite{Hinderer:2007mb,Hinderer:2009ca,Postnikov:2010yn}
\begin{equation}
\Lambda=\frac{2k_2}{3C^5} .
\end{equation}
with $C=M/R$ as the compactness of the star.
The tidal deformability $\Lambda$ is usually solved simultaneously with
the TOV equations. 

\subsection{Chirp Mass}
The chirp mass $\mathcal{M}$ describes the inspiral phase of gravitational wave emission and is given as:
	\begin{equation}
		\mathcal{M}=\left(\frac{q}{(1+q)^2}\right)^{\frac{3}{5}}M_{total}
	\end{equation}
where q is the mass ratio $M_1/M_2$ of the participating compact stars and $M_{total}$ is their total mass. The total mass and chirp mass are, on a qualitative level, easily interchangeable, because the value of the prefactor is not strongly effected by inserting the most extreme cases for $q$ supported by the LIGO data, which are 1.4 and 1.0. Even the most extreme cases provided by our EoSs would not change the prefactor in a significant way. However the chirp mass can be measured more precisely. In the following we will use the very precise measurement of $\mathcal{M}=1.186^{+0.001}_{-0.001}M_\odot$ by LIGO \cite{Abbott:2018wiz} to constrain our calculations. However, it is more useful to consider $M_{total}$ in order to distinguish between possible merger scenarios (see eqs.\eqref{MainEq}), since it is more intuitive.
%=============================================================================
%=========================================================================
 \section{Results}\label{Manwe}
%==========================================================================
\subsection{Possible Combinations for a given total mass }
\begin{figure*}
 	\centering				
 	\includegraphics[width=14cm]{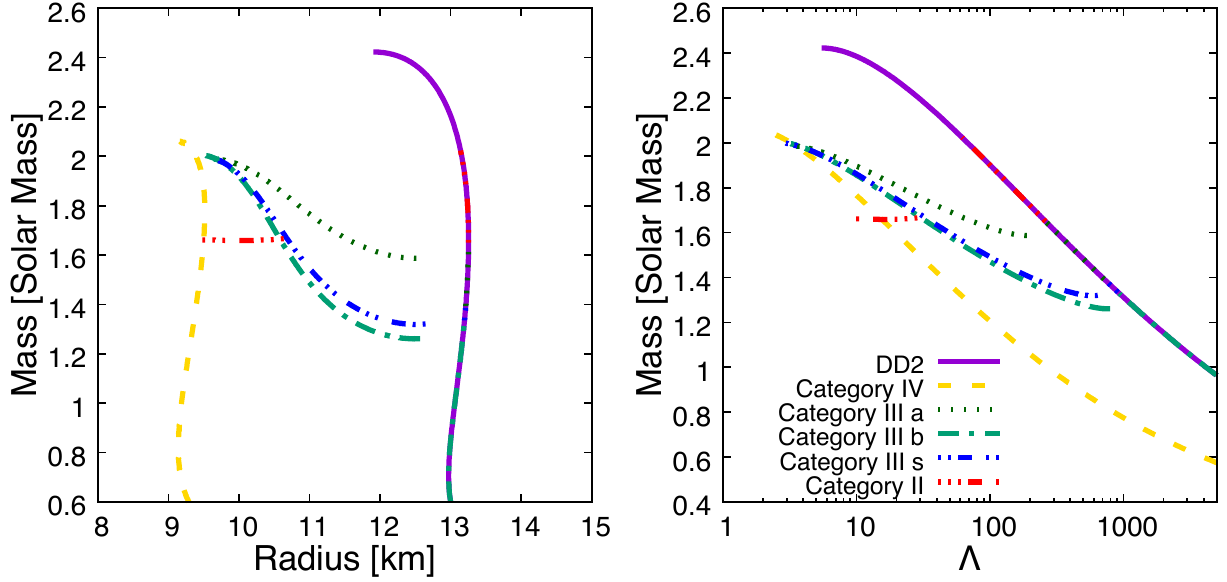}
 	\caption{The left figure shows the mass radius relation of the DD2 EoS and of the categories II, III and IV as discussed in \cite{Christian:2017jni}. The right plot shows the corresponding plot for the mass-$\Lambda$ relation.}
 	\label{MR_Plot}
 \end{figure*}

In this work we explore low-spin star solutions.
When examining the constraints for an EoS in the light of GW170817 one usually considers the tidal deformabilities of both stars participating in the collision and plot them against each other in a $\Lambda_1-\Lambda_2$ plot. Since they are tied together by a total mass $M_{total}$ this plot will result in a broad line, where the position and shape is governed by the EoS and the width by the error of the measurement of $M_{total}$. %\textcolor{blue}{This width can be reduced further by using the chirp mass as a criterion since it can be measured more precicly. However in our following considerations it is favourable to examine the total mass, since it is more intuiviv and all statements would apply in a similar manner to the chirp mass.}\\
The compactness and hence the mass and radius of a star are tightly connected to its tidal deformability (see \cref{MR_Plot}).
%Category I is not included in this article because the masses of its hybrid stars are to high to participate in a BNSM scenario with 2.7$M_\odot$, which means that a category \rom{1} EoS would be identical to the DD2 EoS in a $\Lambda-\Lambda$ plot. The right figure shows the mass as a function of $\Lambda$. The existance of the hybrid star branches, similar to the MR-plot is of particular note.
As a result a twin star pair has two significantly different values of $\Lambda$ as well, where two stars have the same mass but sizably different tidal deformabilities (see Fig. \ref{MR_Plot}). This results in not only one, but two (broad) lines in a $\Lambda_1-\Lambda_2$ plot. This is due to the possibility of pairing a neutron star with another neutron star or a neutron star with a hybrid star and vice versa. The pairings are determined by the total mass, which, for an EoS containing twin stars usually means that one star is from the neutron star branch and one is from the hybrid star branch. It becomes apparent that a general rule predicting how many lines in a $\Lambda_1-\Lambda_2$ plot are to be expected, depending on the total mass $M_{total}$ of the BNSM, can be formulated.\\\\
%First we need to properly describe what the "twin star mass" in this context is. 
%If we consider the set of masses that are part of the neutron star branch ($S_{neutron}$) and the hybrid star branch ($S_{hybid}$) in an EoS, we can discribe any given twin star, of that EoS, as part of the intersecting set of the two ($S_{twin}$). With the understanding of $M_{twin} \in S_{twin}$ it is possible to determine the possible variations of star (i.e. hybrid or neutron star) that took part in the collision depending on the total mass. This behavior can be summarised as follows:\\
When using the term "twin star mass" $M_{twin}$ in the following we are referring to the entire range of mass values that are contained in both branches of the mass-radius plot. The twin star masses of the EoSs used later are listed in \cref{Annatar}. With this understanding of $M_{twin}$ it is possible to determine the possible variations of stars (i.e. hybrid or neutron stars) that take part in the merger depending on the total mass. This can be summarized as follows:
\begin{equation}\label{MainEq}
M_{total}	\begin{cases} < 2M_{twin} \Rightarrow NH, NN \\
						  = 2M_{twin} \Rightarrow HH, NH, NN\\
						  > 2M_{twin} \Rightarrow HH, NH
	\end{cases}
\end{equation}
Where HH, NH and NN denote hybrid-hybrid, neutron-hybrid and neutron-neutron combinations, respectively.\\\\
\subsection{Tidal Deformability with the Total Mass of GW170817}
\begin{figure}
	\centering				
	\includegraphics[width=8.5cm]{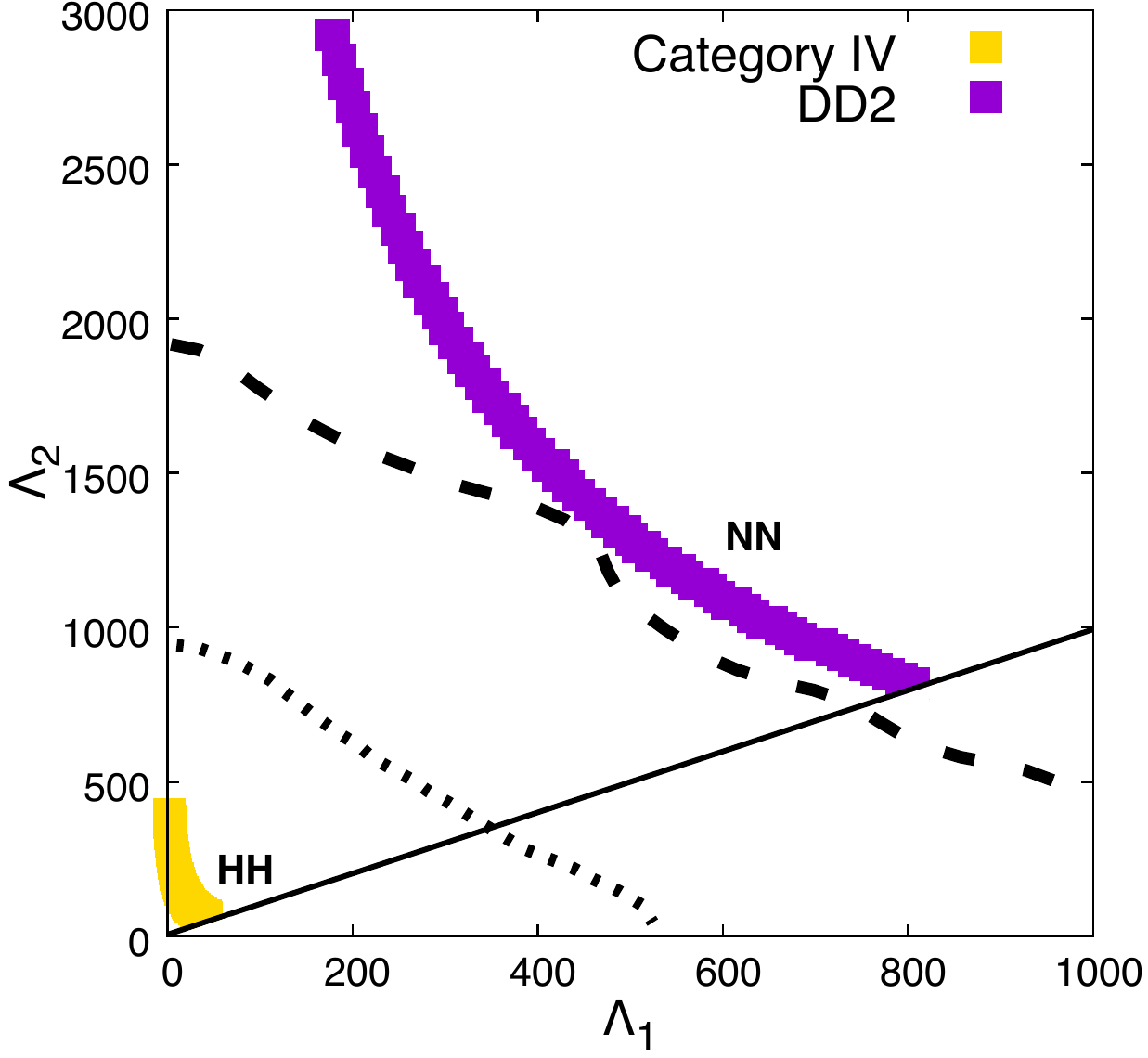}
	\caption{\footnotesize The line created by the DD2 EoS (purple) and the category \rom{4} EoS (gold) with the lowest value of $p_{trans}$ denote the upper and lower limit of the possible $\Lambda-\Lambda$ lines supported in our approach. Neither EoS yields twin stars, which results in only one (broad) line for each case. The black dashed line denotes the 90\% credibility level by LIGO \cite{Abbott:2018exr}, the dotted line describes the 50\% level.}
	\label{LvLCatIVandDD2}
\end{figure}

\begin{figure}
	\centering				
	\includegraphics[width=8.5cm]{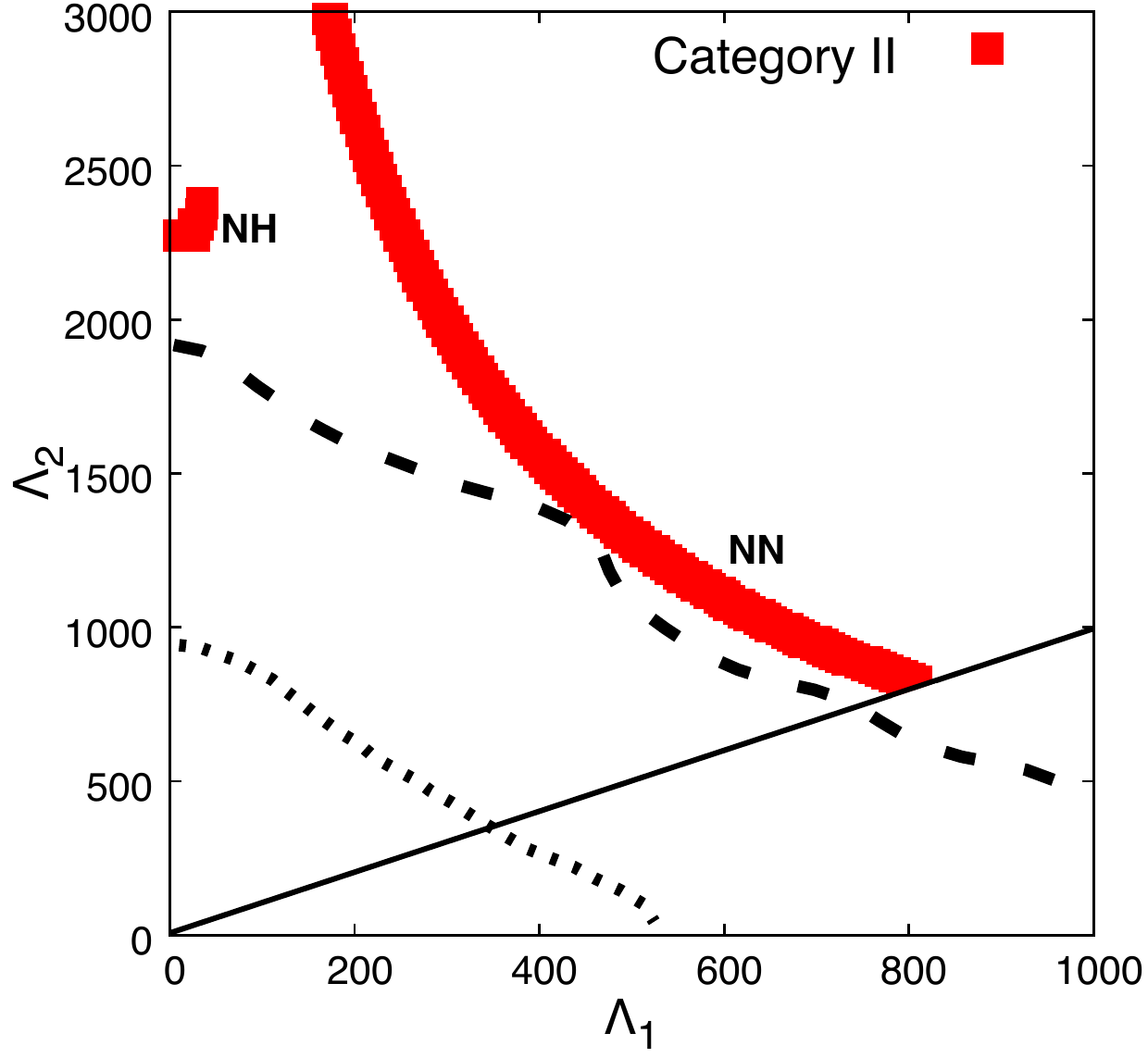}
	\caption{\footnotesize Category \rom{2} results in two lines, of which the upper right is the one generated by the DD2-EoS. The line in the upper left part is a combination of low mass neutron stars with high mass hybrid stars. The values of $\Lambda$ are far too high to be in accordance with GW170817.}
	\label{LvLCatII}
\end{figure}

\begin{figure}
	\centering				
	\includegraphics[width=8.5cm]{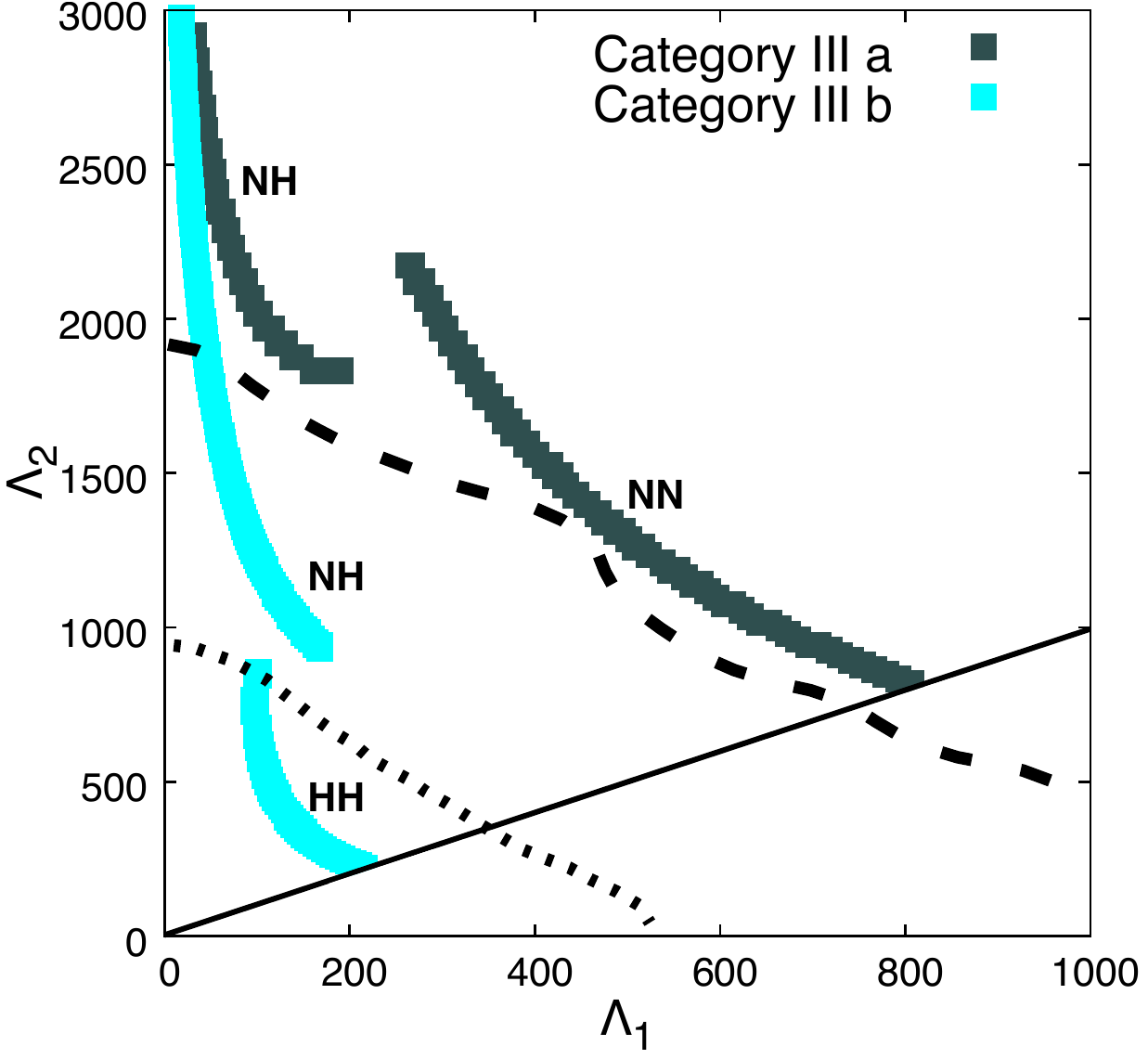}
	\caption{\footnotesize Category \rom{3}a is an example of a category \rom{3} EoS above the $2M_{twin}=M_{total}$ limit. It generates a hybrid-neutron and a neutron-neutron line. Category \rom{3}b is below the limit and generates a hybrid-hybrid and a neutron-hybrid line.}
	\label{LvLCatIIIg}
\end{figure}

In this section we analyse the equations of state of table \ref{Annatar} in light of eq. \eqref{MainEq} using the total mass of ${M_{total}}=2.71^{+0.04}_{-0.01}M_\odot$ measured by LIGO for GW170817 \cite{TheLIGOScientific:2017qsa}. This constraint on the total mass assumes low spin compact stars prior to the GW event. For a rotating star the uncertainty in the total mass increases. However, the figures are constraint using the more precisely determined chirp mass. The qualitative statements are, however, unchanged.\\
Since category \rom{1} contains only hybrid stars with masses of $2M_\odot$ or more its hybrid stars cannot participate in an event with a total mass of $2.7M_\odot$, which makes all $\Lambda_1$-$\Lambda_2$ plots for category \rom{1} identical to the one of the DD2 EoS. This is why no category \rom{1} EoS is included in this work. Category \rom{1} would be only of interest for $M_{total}\ge 4M_\odot$ by definition.\\
The upper and lower limit of the lines found in the $\Lambda_1-\Lambda_2$ plots generated by our parametrization can be seen in Fig. \ref{LvLCatIVandDD2}. The upper limit (purple) is the line generated by the DD2 EoS by \cite{Typel:2009sy}. This is the configuration every neutron-neutron star pair this model leads to. The $\Lambda_1-\Lambda_2$ values of the DD2 EoS are slightly above the credibility levels by LIGO \cite{Abbott:2018exr}, depicted as a black dashed and a black dotted line for the 90\% and 50\% level respectively. However, this circumstance does not change the general behavior explored in this work, which should hold true for differing hadronic EoS. We checked this for a few examples and will report on it in a forthcoming work after more testing.
% ==================================================
% ==================================================
On the other hand the lower limit (gold) in Fig. \ref{LvLCatIVandDD2} given by case CIV features a particularly low transition pressure. This favors compact second branches in the mass-radius relation at low radii \cite{Christian:2017jni}, thus generating small values of $\Lambda$. This lower limit consists entirely of hybrid-hybrid star combinations with the lowest possible compactness in this model. Every other $\Lambda_1-\Lambda_2$ line generated by our parametrization is located between these two limits shown in Fig. \ref{LvLCatIVandDD2}.\\
Since the total mass used here is approximately $2.7M_\odot$ we reach a region, where $2M_{twin} = M_{total}$ holds in category \rom{3}, which encompasses a great variety of mass-radius relations and as a result we find EoSs with $2M_{twin}$ lesser (e.g. CIIIb), greater (e.g. CIIIa) and equal (e.g. CIIIs) to $M_{total}$ in this category alone.\\
Case CII, shown in Fig. \ref{LvLCatII}, and case CIIIa, shown in \cref{LvLCatIIIg}, generate two lines each. The line in the upper right is the DD2 line as shown in Fig. \ref{LvLCatIVandDD2}, while the other line is the corresponding hybrid-neutron star line. 
The possibility of having two different lines in a $\Lambda_1-\Lambda_2$ plot has already been seen in previous works \cite{Paschalidis:2017qmb,Alvarez-Castillo:2018pve,Sieniawska:2018zzj}. 
% ==================================================
% ==================================================
Since category \rom{2} contains only high mass hybrid stars, having low values of $\Lambda$, they need to be combined with low mass neutron stars, having a high value of $\Lambda$. Thus, the thin line in the left upper corner of \cref{LvLCatII} is created. Higher twin star masses would move the line to higher values of $\Lambda_2$.
% ==================================================
% ==================================================

The twin star mass of CIIIa and CII is approximately identical, however the range of possible star combinations becomes larger (compare $M_{twin}$ in \cref{Annatar}) creating a wider line in CIIIa, see \cref{LvLCatIIIg}, due to the more compact second branch in the mass-radius relation, see Fig. \ref{MR_Plot}. One should also note that the neutron-neutron line generated by CIIIa is cut compared to CII, as there are fewer eligible pure neutron star configurations in the EoS.\\

\begin{figure}
	\centering				
	\includegraphics[width=8.5cm]{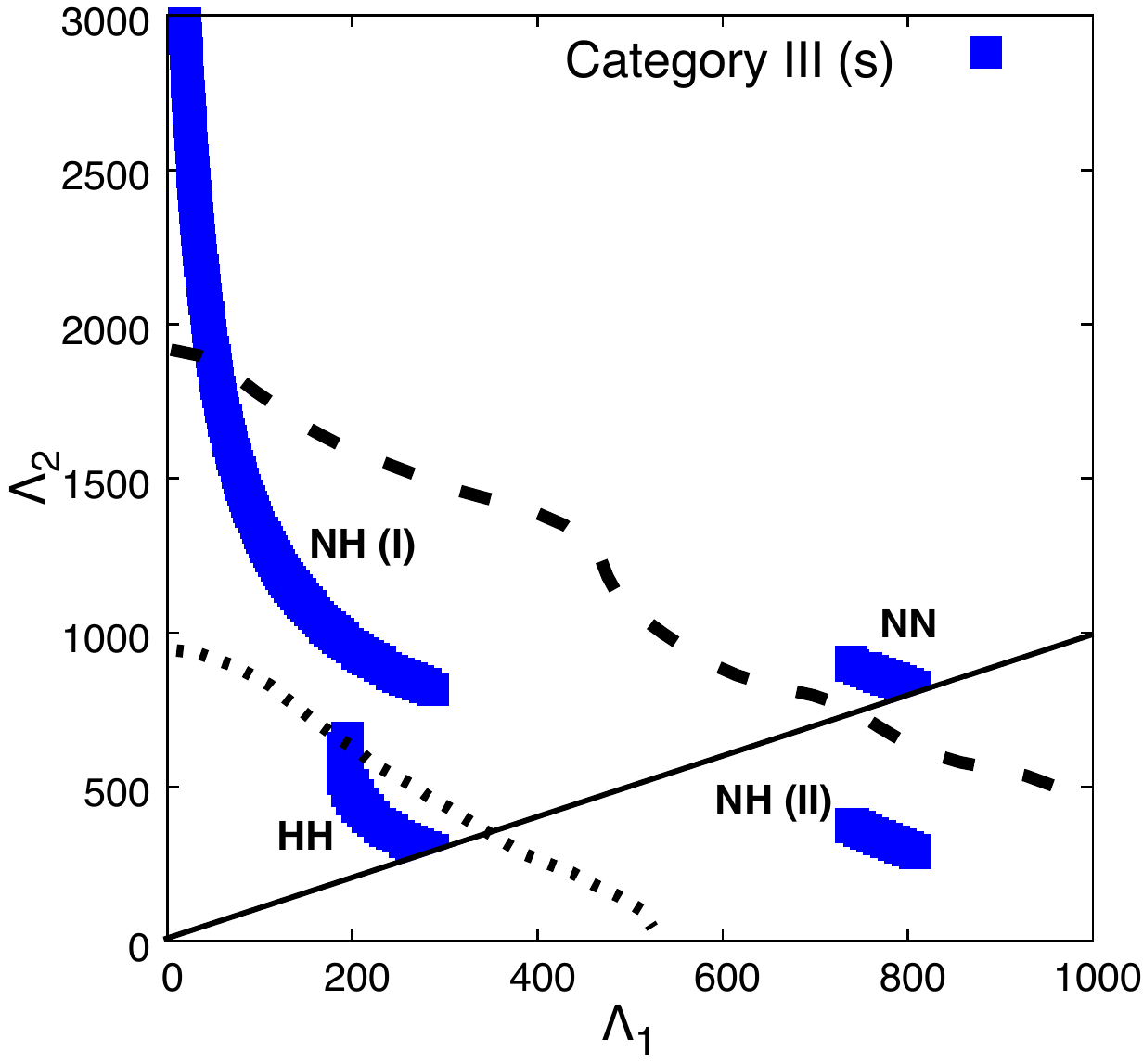}
	\caption{\footnotesize This configuration depicts the special case in which $M_{total} = 2M_{twin}$ is fulfilled. We find hybrid-hybrid, hybrid-neutron and neutron-neutron combinations for this EoS. Though not intuitive the neutron-hybrid star line has an extension in the regime where $\Lambda_1 > \Lambda_2$ which is generated by so called "rising twins", where the heavier twin star has a larger radius.}
	\label{LvLCatIIIs}
\end{figure}

The second EoS shown in \cref{LvLCatIIIg}, CIIIb, has its twin star mass located below $0.5M_{total}$. It is thus impossible to find combinations of pure neutron stars that add up to the total mass. We find a neutron-hybrid star line as well as a hybrid-hybrid star line.  The shape of the hybrid-hybrid star line is very similar to the one of CIV, but the location is at higher values of $\Lambda_1$ and $\Lambda_2$, since the hybrid stars of CIIIb are less compact than the hybrid stars of CIV.\\
Fig. \ref{LvLCatIIIg} suggests that eq.\eqref{MainEq} also gives an indication whether the neutron-hybrid lines generated by a given EoS are located above or below the LIGO limit. 
The remaining $\Lambda_1-\Lambda_2$ diagrams seem to support this observation, where an EoS that generates NH and NN lines (i.e. one with $2M_{twin} > M_{total}$) has its NH line above the 90\% credibility level and EoS that generate NH and HH ($2M_{twin} < M_{total}$) below.
This behavior could be caused, because the eligible stars of the neutron part in the mass-radius relation for a total mass are more numerous, dominating their impact on the locations on the NH line. However, whether this is a coincident or a general rule remains to be determined in future works.
If the total mass equals $2M_{twin}$, for any of the twin stars, we observe four distinct lines in the $\Lambda-\Lambda$ plot. This is depicted in Fig. \ref{LvLCatIIIs}, where a hybrid-hybrid line emerges in addition to the neutron-hybrid and hybrid-hybrid lines of category \rom{2} and category \rom{3}a. 
% =========================================
% =========================================
This demonstrates that generating a hybrid-hybrid, neutron-hybrid and neutron-neutron star line with just one EoS introducing a sharp phase transition is possible as well. The EoS describing this phenomenon is CIIIs. It is only possible to find such a configuration in a very small parameter space. This becomes apparent when comparing with CIIIb, where no neutron-neutron line appears in the $\Lambda-\Lambda$ plot, recall Fig. \ref{LvLCatIIIg}, even though the mass-$\Lambda$ plot for C\rom{3}b is nearly identical to C\rom{3}s, which can be seen in the right figure of Fig. \ref{MR_Plot}.\\
One should also note that CIIIb in comparison with CIIIa and CIIIs demonstrates, how lower values for $M_{twin}$ lead to lines closer to the CIV limit in the $\Lambda-\Lambda$ plot (see Fig. \ref{LvLCatIIIg}).\\
By definition we only plot combinations with $M_1 > M_2$. With this assumption one would expect that $\Lambda_2 >\Lambda_1$ follows. However, there is a configuration where this is not true. This configuration are the so called "rising twins", which are twin star combinations where the more massive one has a larger radius \cite{Schertler:2000xq}, it can be seen in figure \ref{MR_Plot} for the case CIIIs. The "rising twin line" NH(II) in Fig. \ref{LvLCatIIIs} is positioned in the lower right corner and is an extension of the neutron-hybrid line NH(I). The tidal deformability depends on the compactness as $\Lambda \propto 1/C^6$ \cite{Zhao:2018nyf}. In order to find values of $\Lambda_1 >\Lambda_2$ it has to follow, that the compactness $C_1$ is smaller than $C_2$, which is the case when $0 < \Delta M/\Delta R < M_1/R_1$.

% ==================================================
\begin{figure}
	\centering				
	\includegraphics[width=8.5cm]{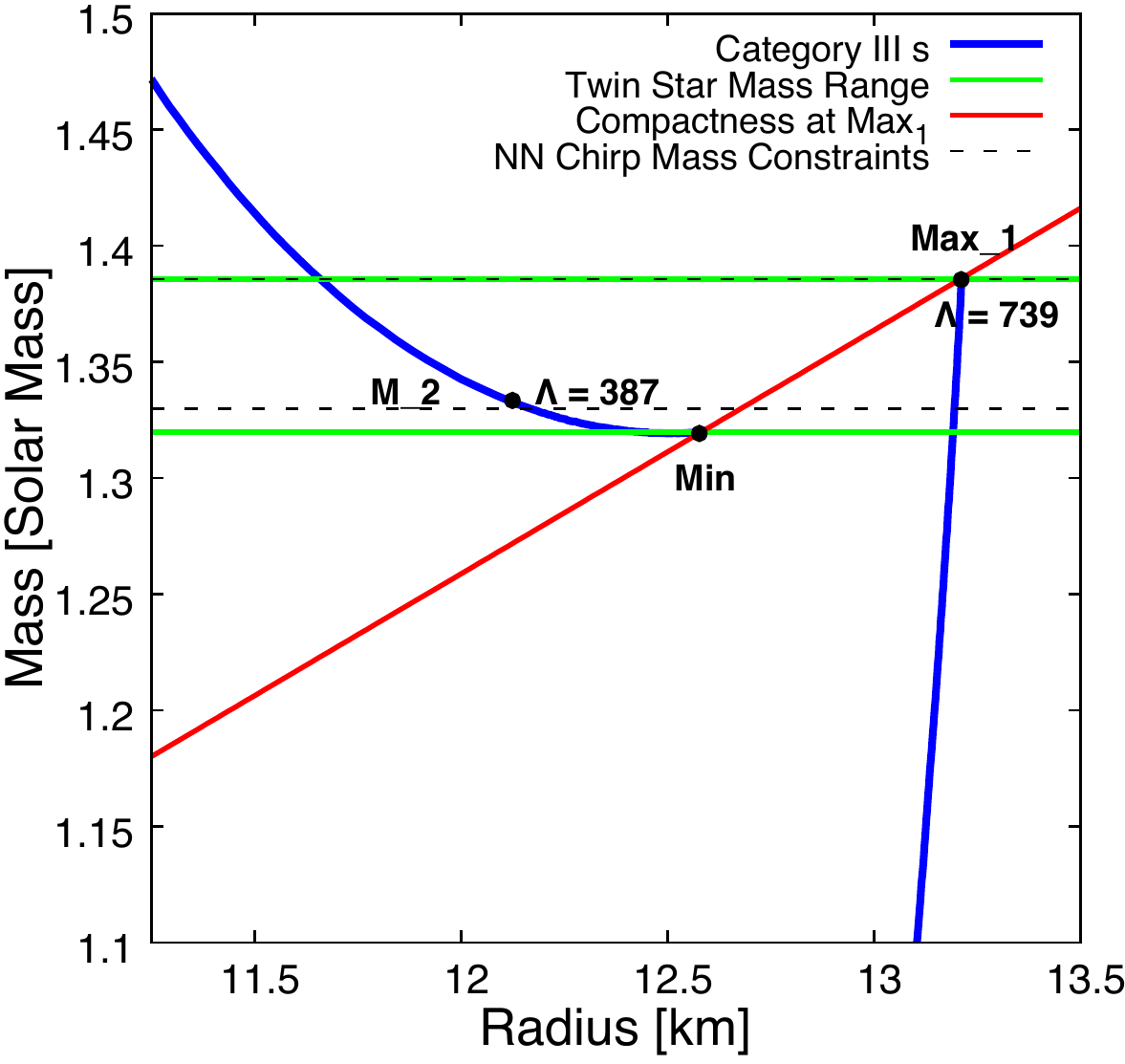}
	\caption{\footnotesize Mass radius relation for category III(s). The green horizontal lines represent the mass at the maximum of the neutron star branch and the minimum mass for the hybrid star branch. The dotted lines are the boundaries given by the LIGO measurements and the red line represents the compactness M/R for $\mathrm{Max_1}$. The point $\mathrm{M_2}$ is an possible companion for $\mathrm{Max_1}$, generating a point in the NH(II) line. %In between the two dotted lines NN solutions are possible and between the green horizontal lines NH(II) solutions are possible. NH(I) solutions can be found for the entire mass range.
	}
	\label{unphys}
\end{figure}
% ==================================================
Fig.\ref{unphys} shows the mass radius relation for category III(s) by the blue lines. 
The green horizontal lines represent the mass at the first maximum $\mathrm{Max_1}$ of the neutron star branch and the mass of the hybrid star branch minimum Min. For any given total mass the green horizontal lines  indicate where NH(II) solutions are supported by this EoS. The dotted lines are the boundaries given by the LIGO measurement for the %total mass $M_{total}$, i.e. the upper line is 2.78$\rm{M_{\odot}}$/2 and the lower line is 2.73$\rm{M_{\odot}}$/2. 
chirp mass under the assumption, that the more massive star does not exceed a mass of $M_{max_1}$. Thus the dotted lines mark the area where NN solutions are found. It follows, that the upper half of this region is where neutron stars that are part of NH(II) solutions originate from, while the lower half provides the corresponding hybrid stars.
The red line represents the compactness M/R at $\mathrm{Max_1}$. Above the red line and below $M_{Max_1}$, hybrid stars are more compact than the neutron star at $M_{Max_1}$ so that $\Lambda_1 > \Lambda_2$. %Between the two dotted lines NN solutions are possible.
The HH line contains stars with masses between $1.32M_\odot$ (the lower green line) and $1.4M_\odot$. The point $\mathrm{M_2}$, together with $\mathrm{Max_1}$, generates an entry in the NH(II) line. There respective values of $\Lambda$ have been added to illustrate that point. NH(II) solutions are the special case where $\Lambda_2<\Lambda_1$, see also figure \ref{LvLCatIIIs}. NH(I) solutions can be found for the entire mass range, as long as $M_1\geq\ M_2\geq 1M_{\odot}$ is satisfied and $M_1$ and $M_2$ add up to a chirp mass of $1.186^{+0.001}_{-0.001}M_\odot$. 
\subsection{Changes in the position of the $\Lambda$-$\Lambda$ line with increasing total mass}
\begin{figure}
	\centering				
	\includegraphics[width=8.5cm]{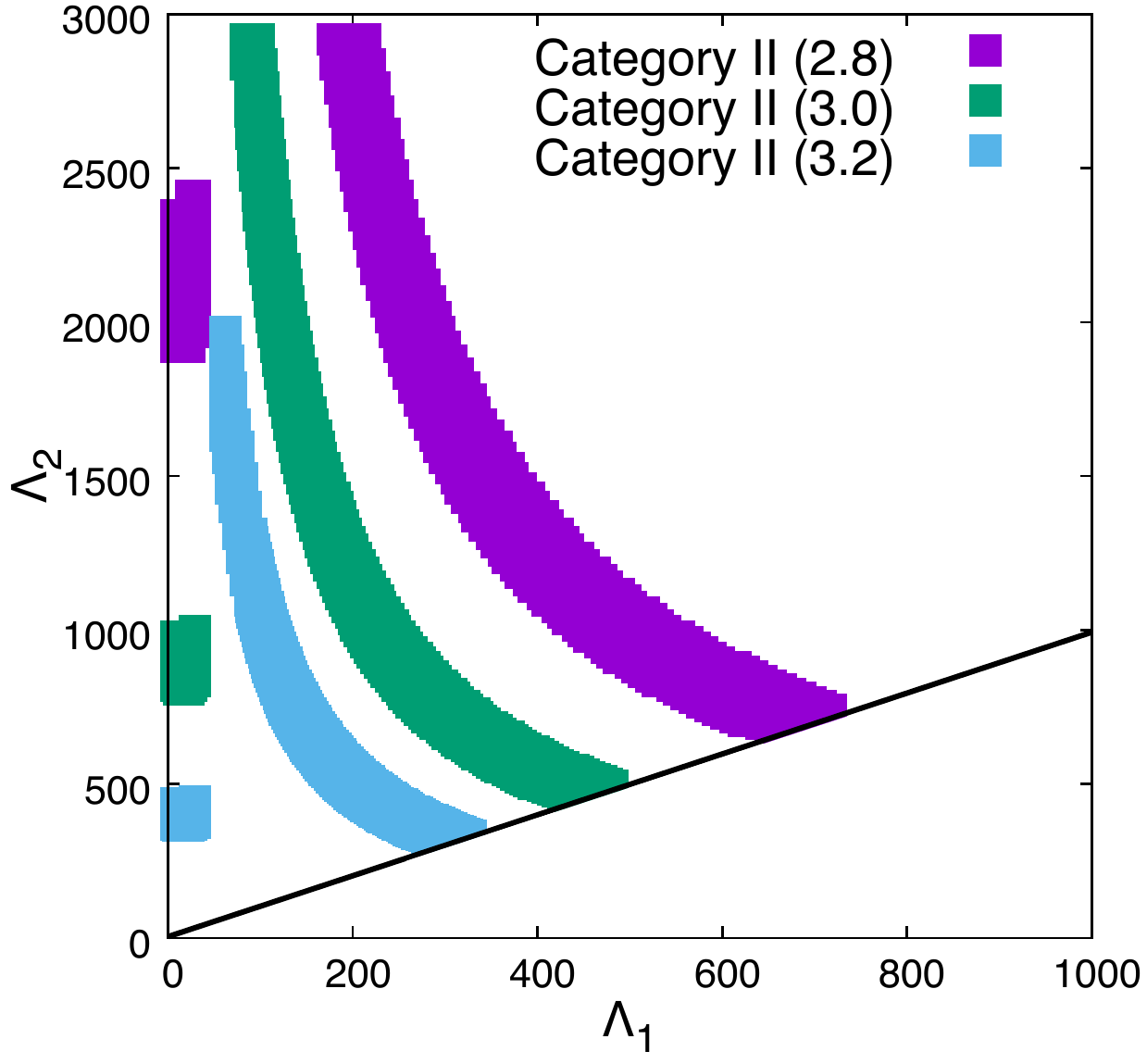}
	\caption{\footnotesize By increasing the total mass of the BNSM, the hybrid-neutron star line moves to smaller values of $\Lambda_2$ at constant $\Lambda_1$. The entire neutron-neutron branch moves to lower values of $\Lambda$ for both stars.}
	\label{AlttotalmassCatII}
\end{figure}

We demonstrate the effect of different total masses in Fig. \ref{AlttotalmassCatII} with the example of CII. Since a large value of a stars mass results in a smaller value of $\Lambda$ it is to be expected that a higher total mass of a BNSM favors smaller values of $\Lambda$. This behavior is observed in Fig. \ref{AlttotalmassCatII} where three different cases for $M_{total}$ are plotted.  
% ====================================
% ====================================
The shape of these two lines remains independent from $M_{total}$, but the relative position changes to smaller values in general. This is because the shape of the $\Lambda-\Lambda$ line is governed by the EoS and not by the total mass.
For demonstration purposes we used $M_{total}$ to constrain the plot, instead of $\mathcal{M}$ as in the previous figures.
   %=========================================================================
   \section{Summary}\label{mithrandir}
  %==========================================================================
We explored the tidal deformabilities of the twin star solutions obtained with constant speed of sound parametrized EoSs as studied in detail in ref. \cite{Christian:2017jni}.
EoSs with a sharp phase transition may yield up to four broad lines in the $\Lambda-\Lambda$ plot, if the EoSs allow for twin star solutions.
The merger of a neutron-neutron, a neutron-hybrid and a hybrid-hybrid star are %hence
possible combinations for GW170817, with a small number of EoSs even allowing for all three combinations at the same time. The combinations supported by an EoS are determined by the total mass of the merger event.
We furthermore show that, when allowing for a larger $M_{total}$ of the BNSM, the $\Lambda-\Lambda$ band shifts to lower values because the merging stars are allowed to have higher masses, resulting in more compact stars.\\\\ 
The measurement of GW170817 has yielded an estimate for the tidal deformability of compact stars. In the first analysis of the LIGO/Virgo collaboration \cite{TheLIGOScientific:2017qsa} the dimensionless tidal deformability of a $1.4 M_{\odot}$ star has to be $\Lambda \leq 800$. An updated version from the LIGO/Virgo collaboration reports a value of $\Lambda_{1.4M_{\odot}}=190^{+390}_{-120}$ at 90$\%$ credible level \cite{Abbott:2018exr}, which implies that rather soft EoSs (APR4) are favored over stiff ones (H4 or MS1). 
Coughlin et al \cite{Coughlin:2018miv} on the other hand constrain the EoSs once information of the electromagnetic radiation is combined with the GW signal. They state that rather soft EoSs are ruled out by introducing a lower limit on the (dimensionless) binary tidal deformability $\tilde{\Lambda} \geq 197$.
Dai et al \cite{Dai:2018dca} however state that $\tilde{\Lambda}$ is strongly effected by the GW frequency cutoff used in the analysis, so that the theoretical uncertainty of the existing waveform templates may play a crucial role in the determination of $\tilde{\Lambda}$.\\
 %%%%%%%%%%%%%%%%%%%%%%%%%%%%%%%%%%%%%%%%%%%%%%%%%%%%%%%%%%%%
Several works have been published in order to constrain the radius of neutron stars \cite{TheLIGOScientific:2017qsa,Annala:2017llu,Abbott:2018exr,De:2018uhw,Tews:2018iwm,Most:2018hfd}. 
In the second version of Ref.~\cite{De:2018uhw} 
$\tilde{\Lambda}=233^{+448}_{-144}$ at 90$\%$ credible level for a component mass prior informed by radio pulsars implying $R_{1.4 M_{\odot}} \simeq 10.8^{+2.1}_{-1.6}\pm 0.2$~km. 
LIGO/Virgo \cite{Abbott:2018exr} reports $10.5\leq R_{1.4M_{\odot}}({\rm km})\leq 13.3$ for EoSs which support maximum masses of $M \geq 1.97 M_{\odot}$.
Reference \cite{Annala:2017llu} conclude that the maximum radius of a 1.4$M_{\odot}$ star is 13.4~km for a tidal deformability of $\Lambda=224$ and in 
Ref. \cite{Tews:2018iwm} a similar value is found, where $R_{1.4M_{\odot}} \leq 13.6$~km. 
Most et al. \cite{Most:2018hfd} constrain the radius to a smaller area $12.00\leq R_{1.4M_{\odot}}({\rm km})\leq 13.45$ with a 2-$\sigma$ confidence level for neutron stars, where the lower limit is a result of the constraints put on $\Lambda$ by the afterglow derived by Radice et al. \cite{Radice:2017lry}. 
Allowing for a phase transition, i.e. twin star branches, they find even smaller radii $8.53\leq R_{1.4M_{\odot}}({\rm km})\leq 13.74$.
A phase transition weakens the correlation of the tidal deformability of the neutron star merger \cite{Zhao:2018nyf}, i.e. if one assumes that compact stars are not necessarily categorized in just one family, 
one can indeed explain very small radii \cite{Burgio:2018yix}. 
All these results however are in a good agreement with our results concerning the radius of a 
$1.4M_{\odot}$ compact star, see fig. \ref{MR_Plot} and for more details see ref. \cite{Christian:2017jni}.\\
Alvarez-Castillo et al. \cite{Alvarez-Castillo:2018pve} construct twin star solutions within a nonlocal chiral quark matter EoS for the stars core. In their work they demonstrate that GW170817 may be interpreted as a merger of a neutron-hybrid or a hybrid-hybrid star, their model however does not include the case of neutron-neutron star mergers.  
The possibility of a phase transition in dense matter has also been discussed by Paschalidis et al. \cite{Paschalidis:2017qmb}. Based on a parametrized hybrid hadron EoSs, they utilize  a similar EoS for the dense core as in our approach.
They find that a sufficiently stiff hadronic EoS may be inconsistent with GW170817, but that a hadron-quark phase transition in the compact star can soften the
EoS to make it compatible with GW170817. Most important to note is, that they find that  GW170817 is entirely consistent with merger
of a hybrid star with a neutron star, which is also in accordance with our findings. However, their model only allows for neutron-neutron and hybrid-neutron merger combinations.\\  
Reference \cite{Sieniawska:2018zzj} compares tidal deformabilities of stars with
weak and strong phase transitions. They find constraints on the EoSs and state that most of the high-mass twins can be formed for minimal values of the density jump and that the minimal radius on the twin branch lies in between 9.5 and 10.5~km.
Reference \cite{Han:2018mtj} also investigates phase transitions. In accordance with our results, they find that sharp phase transitions lead to the smallest possible tidal deformabilities.\\

In summary, we find that  a phase transition in compact stars is not ruled out by the tidal deformability constraint from GW170817. GW170817 can hence be interpreted as a merger scenario of neutron-neutron, neutron-hybrid or a hybrid-hybrid star where even all three merging scenarios can potentially be present for just one EoS.
Future GW detections of neutron star mergers for $M_{total}\geq 2.7M_{\odot}$ are expected to see even lower values of $\Lambda$, and are expected to tighten the constraints on the EoSs. Having measured several pairs of $\Lambda_1$ and $\Lambda_2$ in future gravitational wave detections, 
the same value for
$\Lambda_1$ but different values for $\Lambda_2$ indicate 
the existence of a strong first order phase transition in dense 
matter as potentially present in compact stars. \\
Space missions such as NICER \cite{2014SPIE.9144E..20A} will be able to offer precise measurements
of masses and radii in the near future. Together with future GW detections the possible existence of a phase transition could well be put under a stringent test.
  %=============================================================================
\begin{acknowledgements}
AZ and JS acknowledge support from the Helmholtz International Center for FAIR (HIC for FAIR). 
\end{acknowledgements}
%===============================================================================
\bibliographystyle{apsrev4-1}
\bibliography{neue_bib}
\end{document}